\documentclass[letterpaper, 10 pt, conference]{ieeeconf}  

\IEEEoverridecommandlockouts                              

\usepackage{amsmath}
\usepackage{amssymb}
\usepackage{graphicx}
\usepackage{subcaption}
\usepackage{xcolor}
\definecolor{darkgreen}{RGB}{0,100,0}
\usepackage{algorithm}
\usepackage{url}
\usepackage{algpseudocode}
\usepackage{cite}

\newtheorem{definition}{Definition}
\newtheorem{theorem}{Theorem}
\newtheorem{lemma}{Lemma}
\newtheorem{problem}{Problem}

\newcommand{\ignore}[1]{}

\title{\LARGE \bf
Verification and Synthesis of Compatible Control Lyapunov and Control Barrier Functions
}

\author{
Hongkai Dai$^{*1}$, Chuanrui Jiang$^{*2}$, Hongchao Zhang$^{2}$, and Andrew Clark$^{2}$
\thanks{$^{1}$Hongkai Dai is with the Toyota Research Institute, Los Altos, CA, USA. Email: 
        {\tt\small hongkai.dai@tri.global}}
\thanks{$^{2}$Chuanrui Jiang, Hongchao Zhang and Andrew Clark are with the Electrical and Systems Engineering Department, McKelvey School of Engineering, Washington University in St. Louis, St. Louis, MO 63130
        {\tt\small \{chuanrui@wustl.edu, hongchao, andrewclark\}@wustl.edu}}
        \thanks{$^*$ denotes equal contribution}%
        \thanks{This work was supported by the  Air Force Office of Scientific Research
grants  FA9550-22-1-0054 and FA9550-23-1-0208 and NSF grant CNS-1941670.}%
}

\begin{document}
\maketitle
\thispagestyle{empty}
\pagestyle{empty}

\begin{abstract}

Safety and stability are essential properties of control systems. Control Barrier Functions (CBFs) and Control Lyapunov Functions (CLFs) are powerful tools to ensure safety and stability respectively. However, previous approaches typically verify and synthesize the CBFs and CLFs separately, satisfying their respective constraints, without proving that the CBFs and CLFs are compatible with each other, namely at every state, there exists control actions within the input limits that satisfy both the CBF and CLF constraints simultaneously. Ignoring the compatibility criteria might cause the CLF-CBF-QP controller to fail at runtime. There exists some recent works that synthesized compatible CLF and CBF, but relying on nominal polynomial or rational controllers, which is just a sufficient but not necessary condition for compatibility. In this work, we investigate verification and synthesis of compatible CBF and CLF independent from any nominal controllers. We derive exact necessary and sufficient conditions for compatibility, and further formulate Sum-Of-Squares programs for the compatibility verification.
 Based on our verification framework, we also design a nominal-controller-free synthesis method, which can effectively expands the compatible region, in which the system is guaranteed to be both safe and stable. We evaluate our method on a non-linear toy problem, and also a 3D quadrotor to demonstrate its scalability. The code is open-sourced at \url{https://github.com/hongkai-dai/compatible_clf_cbf}.
\end{abstract}

\section{Introduction}
\label{sec: intro}

Ensuring safety and stability are essential for control systems to prevent catastrophic economic damage and loss of human life, while still achieving the desired goal \cite{dhscps, li2023survey, fachri2022multiple, cohen2024safety}. Lyapunov stability guarantees that all trajectories starting within the system's region-of-attraction (RoA) will converge to the goal state. Meanwhile, safety properties are normally formulated as the positive invariance \cite{ames2014control} of given regions in the state space. 

The importance of safety has motivated numerous methods such as artificial potential fields \cite{fan2020improved}, reachability analysis\cite{chen2022reachability}, Hamilton--Jacobi analysis\cite{lee2020hopf}, model predictive control \cite{schwenzer2021review}, and control barrier functions \cite{ames2019control} to synthesize safe control strategies.
Among these methods, energy-based methodologies such as barrier certificates \cite{prajna2007framework} and Control Barrier Functions (CBFs) \cite{ames2019control} have been proposed that construct an energy function that is positive if the system is safe, and then demonstrate safety by proving that the energy function remains positive for all time. 
Similarly, Lyapunov stability is certified by a Lyapunov function \cite{lyapunov1992general} or a Control Lyapunov function (CLF) \cite{sontag1983lyapunov}, which remains non-negative and decreasing along any trajectories, and achieves the minimal value at the goal state.
One major benefit of CLFs/CBFs is that they can be incorporated as constraints within optimization-based controllers, resulting in CLF-CBF control policies for joint safety and stability \cite{ames2014control, zeng2021safety}. 
Such techniques have been widely used in  safety critical scenarios such as adaptive cruise control\cite{ames2014control, cohen2022high}, bipedal robot control \cite{ames2019control, choi2020reinforcement}, and multi-waypoint navigation \cite{fachri2022multiple}, with many different approaches to synthesize the CLF and CBF functions \cite{tan2004searching, dawson2023safe, liu2023safe, wang2023safety, kang2023verification, dai2023convex, clark2022semi, zhao2023safety, chen2024learning, zhang2023efficient, yang2024learning, dai2022learning, jin2020neural, zhao2023convex, wang2024}.

\begin{figure}
\includegraphics[width=0.45\textwidth]{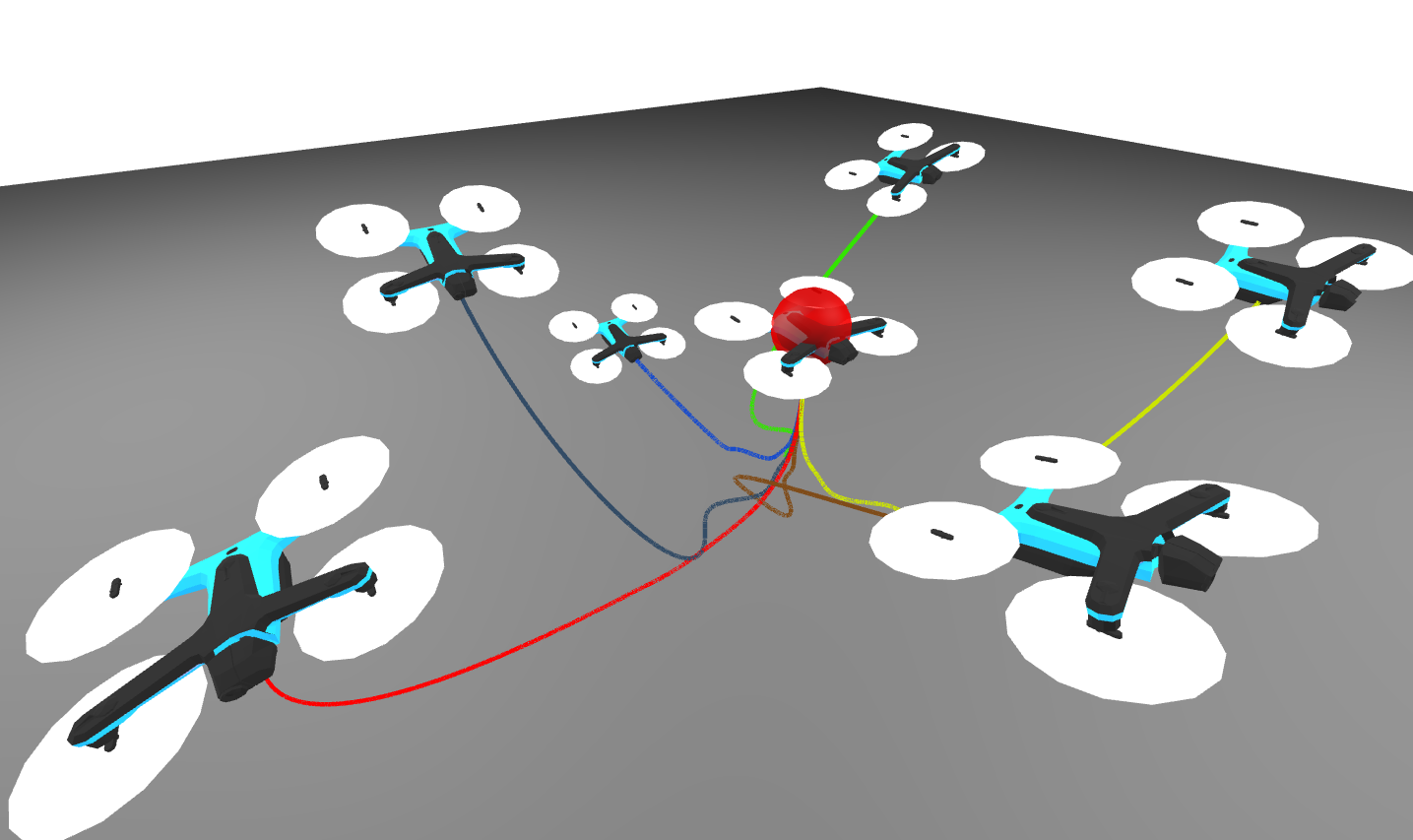}
\caption{With the CLF-CBF-QP controller, the quadrotor converges to the desired goal (red sphere in the center) from different initial states, while avoiding the grey ground. The compatible CLF/CBF are synthesized by our algorithm. We also plot each trajectory of the quadrotor body origin with a different color.}
\label{fig:quadrotors}
\end{figure}
A key challenge in CLF-CBF-based safe control policies is the \emph{compatibility} of the CLF and CBF constraints, namely the existence of at least one admissible control action satisfying both CLF and CBF constraints simultaneously at every state in the safe and stabilizable region. Most previous approaches address incompatibility by relaxing the CLF constraints \cite{ames2014control}, and thus prioritizing safety over stability. However, this may sacrifice the stability and result in the system remaining in undesired states \cite{reis2020control}. Recent efforts \cite{schneeberger2024advanced, schneeberger2023sos, isaly2024feasibility} have utilized Sum-Of-Squares (SOS) methods \cite{papachristodoulou2013sostools, parrilo2000structured} to verify and synthesize  compatible CLF and CBFs based on nominal controllers, for continuous-time control-affine systems with polynomial dynamics. In such methods, a nominal controller is parameterized as a polynomial (or rational) function of state; these methods search for the CLF/CBF together with this nominal controller, such that this nominal controller simultaneously satisfies both the CLF and CBF constraints. The disadvantage of this approach is that it is only a sufficient but not necessary condition for compatibility, as there might exist a non-polynomial (or non-rational) compatible controller, which cannot be captured by the explicit parameterization\ignore{, and 2) This approach leads to an expensive tri-linear alternation process, which alternates between fixing the controller, the CLF/CBF functions, and the Lagrangian multipliers}. A related but orthogonal direction of research synthesizes closed-form controllers for CBF and CLF that are known to be compatible \cite{li2023graphical}.


In this paper, we study the problem of exact verification and synthesis of compatible CLF/CBF for all states in the safe region.
We consider a nonlinear control-affine system, both with and without input limits, and derive  necessary and sufficient conditions for  compatibility using Farkas' Lemma and the algebraic-geometric Positivstellensatz. 
We formulate the exact verification as a SOS program and further simplify it using S-procedure. 
To synthesize compatible CLF/CBF,  we propose a bi-linear alternating method based on the simplified verification program. 
To summarize, this paper makes the following contributions:
\begin{itemize}
    \item We derive the necessary and sufficient condition for CLF-CBF compatibility, and formulate a SOS program independent from any nominal controllers for compatibility verification.
    \item We next propose an alternating method based on verification program to synthesis compatible CBF and CLF, also independent from any nominal controllers.
    \item We demonstrate our method on a toy system and showcase its scalability on a 13-state quadrotor. 
\end{itemize}

The rest of the paper is organized as follows. 
Section \ref{sec:mathematical_background} introduces the mathematical background on Positivstellensatz and Farkas' lemma.
Section \ref{sec: prelim} presents the definition of the system model, CBF and CLF. The proposed verification and synthesis framework are introduced in Section \ref{sec: methodology}. Section \ref{sec: results} gives the simulation results, and we conclude the paper in Section \ref{sec: conclusion}.

\textbf{Notation}: We use lower case letter ($x$) to denote a scalar, bold lower case ($\mathbf{x}$) to denote a vector, upper case letter ($X$) for a matrix, and calligraphic letter ($\mathcal{X}$) to denote a set.
\section{Mathematical background}
\label{sec:mathematical_background}
A polynomial $p(\mathbf{x})$ is sum-of-squares (sos) iff
\begin{equation}
    p(\mathbf{x}) = \sum_{i=1}^{n} (p_i(\mathbf{x}))^2 \nonumber
\end{equation}
for some polynomials $p_i(\mathbf{x})$. A sos polynomial is non-negative by definition. We can verify a polynomial being sos through convex optimization.

The cone generated from a set of polynomials $\phi_{1},\ldots,\phi_{s}$ is given by 
\begin{multline*}
\Sigma[\phi_{1},\ldots,\phi_{s}] = \left\{\sum_{S \subseteq \{1,\ldots,s\}}{\alpha_{S}(\mathbf{x})\prod_{i \in S}{\phi_{i}(\mathbf{x})}} : \right. \\
\left.\alpha_{S}(\mathbf{x}) \in \mbox{ SOS } \forall S \subseteq \{1,\ldots,s\}\right\}.
\end{multline*}

The ideal generated from polynomials $\psi_{1},\ldots,\psi_{r}$ is given by 
\begin{multline*}
\mathbb{I}[\psi_{1},\ldots,\psi_{r}] \\
= \left\{\sum_{l=1}^{r}{a_{i}(\mathbf{x})\psi_{l}(\mathbf{x})} : a_{1},\ldots,a_{r} \mbox{ polynomials}\right\}.
\end{multline*}

\begin{theorem}[Positivstellensatz (P-satz)\cite{parrilo2000structured}] 
\label{th:Positivstellensatz}
Let $\left(\phi_{j}\right)_{j=1, \ldots, s}$, $\left(\psi_{\ell}\right)_{\ell=1, \ldots, r}$ be finite families of polynomials in $\mathbb{R}\left[\mathbf{x}\right]$. Denote the cone generated from $\left(\phi_{j}\right)_{j=1, \ldots, s}$ by $\Sigma[\phi_{1},\ldots,\phi_{s}]$ , and the ideal generated from $\left(\psi_{\ell}\right)_{\ell=1, \ldots, r}$ by $\mathbb{I}[\psi_{1},\ldots,\psi_{r}]$. Then, the following statements are equivalent:
\begin{enumerate}
    \item The set
    $$
    \left\{
    \begin{array}{l|ll}
    \mathbf{x} \in \mathbb{R}^{n} &
    \begin{array}{ll}
    \phi_{j}(\mathbf{x}) \geq 0, & j=1, \ldots, s \\
    \psi_{\ell}(\mathbf{x})=0, & l=1, \ldots, r
    \end{array}
    \end{array}
    \right\}
    $$
    is empty.
    \item There exist $\theta(\mathbf{x}) \in \Sigma, \lambda(\mathbf{x}) \in \mathbb{I}$ such that $\lambda(\mathbf{x})+\theta(\mathbf{x}) + 1 = 0$.
\end{enumerate}
\end{theorem}
In Positivstellensatz, (2) is a necessary and sufficient condition for (1). On the other hand, (2) requires computing the cone $\Sigma$, which needs exponential number of polynomials $\alpha_S(\mathbf{x})$ (as $S$ permutes over all subsets of $\{1,\hdots, s\}$). Besides, the ``necessary" part is guaranteed only when the degrees of $\alpha_S(\mathbf{x})$ in the cone $\Sigma[\phi_1,\hdots,\phi_s]$ and $a_i(\mathbf{x})$ in the ideal $\mathbb{I}[\psi_1,\hdots,\psi_r]$ are sufficiently high, leading to excessively large optimization programs \cite{nie2007complexity}. To remedy this computation challenge, people often resort to \textit{S-procedure}\cite{parrilo2000structured}, which replaces the cone $\Sigma$ with the \textit{quadratic module} $\mathcal{Q}$
\begin{multline*}
    \mathcal{Q}[\phi_1,\hdots, \phi_s] = \\\{\alpha_0(\mathbf{x}) + \alpha_1(\mathbf{x})\phi_1(\mathbf{x}) + \hdots + \alpha_s(\mathbf{x})\phi_s(\mathbf{x}), \alpha_i(\mathbf{x}) \in \text{SOS}\}.
\end{multline*}
\begin{lemma}[S-procedure]
\label{lemma:s-procedure}
    Statement (2) in theorem \ref{th:Positivstellensatz} is a sufficient condition for (1) if we replace the cone $\theta\in\Sigma$ with the quadratic module $\theta\in\mathcal{Q}[\phi_1,\hdots,\phi_s]$.\footnote{Under some stronger assumptions on $\phi(\mathbf{x}), \psi(\mathbf{x})$, such as \textit{Archimedean}, S-procedure also gives a necessary and sufficient condition \cite{blekherman2012semidefinite}.}
\end{lemma}

To reduce the computation burden, in the \textit{S-procedure}, we typically restrict both $\alpha_i(\mathbf{x})$ in the quadratic module, and $a_i(\mathbf{x})$ in the ideal to relatively low-degree polynomials.

Farkas' Lemma gives conditions for non-existence of solutions to linear inequalities and is presented as follows.

\begin{lemma}[Farkas' Lemma \cite{matousek2006understanding}]
\label{Lemma:Farkas}
Let $\Lambda\in\mathbf{R}^{m\times n}$ be a matrix, and $\mathbf{\xi} \in \mathbb{R}^{m}$ be a vector. Then the system of inequalities $\Lambda {\mathbf{x}} \leq \mathbf{\xi}$ has a solution if and only if 
the set $\{\mathbf{z}\;|\; \mathbf{z} \geq 0, \mathbf{z}^{T} \Lambda=\mathbf{0}^{T}, \mathbf{\xi}^T\mathbf{z} =-1\}$ is empty. 
\end{lemma}

\section{Problem statement}
\label{sec: prelim}
\subsection{System Model}
We consider a continuous-time nonlinear control-affine system in the form
\begin{subequations}
\begin{align}
    \Dot{\mathbf{x}} = \mathbf{f}(\mathbf{x}) + g(\mathbf{x})\mathbf{u}\\
    \mathbf{u}\in\mathcal{U}=\{ \mathbf{u}\;|\; A\mathbf{u}\le \mathbf{c}\},
\end{align} 
\label{eq:sys_dyn}

where $\mathbf{x}\in \mathcal{X} \subset \mathbb{R}^{n_x}$ is the state and $\mathbf{u} \in \mathcal{U} \subset \mathbb{R}^{n_u}$ is the control input. We also assume that $f:\mathcal{X} \rightarrow \mathbb{R}^{n_x}$ and $g:\mathcal{X} \rightarrow \mathbb{R}^{n_x\times n_u}$ are polynomial functions of $\mathbf{x}$. The set of admissible control $\mathcal{U}$ is represented as a polyhedron $A\mathbf{u}\le \mathbf{c}, A\in\mathbb{R}^{p\times n_u}, \mathbf{c}\in\mathbb{R}^{p}$. Note that this parameterization of $\mathcal{U}$ includes the entire set $\mathcal{R}^{n_u}$ (unconstrained control) as a special case, where $A$ and $\mathbf{c}$ are empty, hence we can model systems both with and without input limits. Without loss of generality we assume that the equilibrium state is $\mathbf{x}=\mathbf{0}$. 

Many interesting dynamical systems don't exhibit polynomial dynamics at  first glance (namely $f(\mathbf{x}), g(\mathbf{x})$ are not polynomials). For example, robots with revolute joints typically use joint angle $\theta$ as the state, and its dynamics involves $\sin\theta$ and $\cos\theta$ functions, which are not polynomials. To overcome this, a widely-used technique is a change-of-variable trick: instead of using $\theta, \dot{\theta}$ as the state, we use $s = \sin\theta, c=\cos\theta$ and $\dot{\theta}$ as the state, with the time derivative as $\dot{s} = \dot{\theta}c, \dot{c}=-\dot{\theta}s$ both as polynomial functions of the new state $(s, c, \dot{\theta})$. Additionally we need to introduce the constraint $s^2 + c^2 = 1$ on the state. As a result, we also include the algebraic equality constraint
\begin{align}
    e(\mathbf{x}) = 0\label{eq:system_algebraic_constraint}
\end{align}
\end{subequations}
in the system dynamics (where $e(\mathbf{x}) = s^2 + c^2 -1$ for the revolute joint example). $e(\mathbf{x})$ is a polynomial function of $\mathbf{x}$. This technique on converting the non-polynomial dynamics to polynomial one has been used in previous SOS-based controller design\cite{papachristodoulou2002construction, posa2015stability}.

The state space $\mathcal{X}$ is divided into a safe region $\mathcal{X}_s$ and an unsafe region $\mathcal{X}_u = \mathcal{X} - \mathcal{X}_s$. Our goal is to formally certify that within the safe region $\mathcal{X}_s$, a control-invariant set such that all trajectories starting within this control-invariant set will converge to $\mathbf{x} = \mathbf{0}$.

\subsection{Control Barrier Function (CBF) and Control Lyapunov Function (CLF)}
We consider safety constraints that are specified as super-level sets of differentiable functions $h: \mathcal{X} \rightarrow \mathbb{R}$. In particular, we consider a set $\mathcal{C}$ parameterized as 
\begin{align*}
    \mathcal{C} &= \{\mathbf{x} \;|\; h(\mathbf{x}) \geq 0\}.
\end{align*}
If $\mathcal{C} \subseteq \mathcal{X}_{s}$ and $\mathcal{C}$ is control-invariant, then starting from any $\mathbf{x}(0)\in\mathcal{C}$, there exists control actions to keep the trajectory safe for all time $t\ge 0$.
\begin{definition}[CBF \cite{ames2019control}]
\label{def:cbf}
    A function $h(\mathbf{x})$ is a CBF if 
    \begin{equation}
    \label{eq:def_cbf}
        \sup_{\mathbf{u}\in \mathcal{U}}\{L_fh(\mathbf{x}) + L_gh(\mathbf{x})\mathbf{u} + \kappa_h(h(\mathbf{x}))\} \geq 0
    \end{equation}
    for all $\mathbf{x} \in \mathcal{C}$, where $L_fh(\mathbf{x}) = \frac{\partial h(\mathbf{x})}{\partial \mathbf{x}}\mathbf{f}(\mathbf{x})$, $L_gh(\mathbf{x}) = \frac{\partial h(\mathbf{x})}{\partial \mathbf{x}}g(\mathbf{x})$, and $\kappa_h(\cdot)$ is an extended class $\kappa$ function\footnote{Recall function $\kappa(.)$ is extended class $\kappa$ if it is monotone, increasing, and satisfies $\kappa(0)=0$.}. 
\end{definition}
Constraint \eqref{eq:def_cbf} ensures that the set $\mathcal{C}$ is control-invariant for systems represented by \eqref{eq:sys_dyn} \cite{ames2019control}.
    
A control Lyapunov function is defined as follows.

\begin{definition}[CLF \cite{sontag1989universal}]
\label{def:clf}
    A differentiable function $V: \mathcal{X} \rightarrow \mathbb{R}$ is a Control Lyapunov Function if $V(\mathbf{0}) = 0$, $V(\mathbf{x}) > 0 \text{ if } \mathbf{x}\neq 0$ and 
    \begin{equation}
    \label{eq:def_clf}
        \inf_{\mathbf{u}\in \mathcal{U}} \{L_fV(\mathbf{x}) + L_gV(\mathbf{x})\mathbf{u} + \kappa_V(V(\mathbf{x}))\} \leq 0
    \end{equation}
    for all $\mathcal{D}=\{\mathbf{x}\;|\;V(\mathbf{x})\le 1\}$, where $\kappa_V(.)$ is also an extended class $\kappa$ function. This proves that the sub-level set $\{\mathbf{x}\; |\; V(\mathbf{x})\le 1\}$ is an inner approximation of the region-of-attraction (ROA).
\end{definition}

To use our SOS tool, we assume that both the CLF $V(\mathbf{x})$ and CBF $h(\mathbf{x})$ are polynomial functions of $\mathbf{x}$.
We also assume that $\kappa_h(\cdot), \kappa_V(\cdot)$ are polynomial. We will show that for a given pair of CBF/CLF candidates, we can certify the compatibility condition for general polynomial function $\kappa_h(\cdot), \kappa_V(\cdot)$. On the other hand, to synthesize the CBF/CLF, we consider a special class of Definition \ref{def:cbf} and \ref{def:clf} consists of Exponential Control Barrier/Lyapunov Functions, for which $\kappa_h(h(\mathbf{x})) = \kappa_h h(\mathbf{x}), \kappa_V(V(\mathbf{x})) = \kappa_VV(\mathbf{x})$ where $\kappa_h, \kappa_V$ are given positive constants.

\section{Methodology}
\label{sec: methodology}
In this section, we first present our algorithm to verify compatibility of CLF and CBF through convex optimization (section \ref{subsec: verification}); then build upon the verification formulation, we present our algorithm to synthesize compatible CLF/CBF through solving a sequence of convex optimization problems (section \ref{subsec: synthesis}).
\subsection{Verification of compatibility}
\label{subsec: verification}
For a candidate CBF $h(\mathbf{x})$ and a candidate CLF $V(\mathbf{x})$, define 0-super-level set $\mathcal{C}=\{\mathbf{x}\;|\; h(\mathbf{x})\geq 0\}$ and the 1-sub-level set $\mathcal{D}:=\{\mathbf{x}\;|\; V(\mathbf{x})\leq 1\}$. We formally state the compatible CLF-CBF as follows
\begin{problem}[Compatible CLF-CBF]
\label{prob:verification}
    Given extended class $\kappa$ functions $\kappa_V(\cdot)$, $\kappa_h(\cdot)$, a positive definite function $V(\mathbf{x})$, and a function $h(\mathbf{x})$ such that  $\mathcal{C}\subseteq \mathcal{X}_s$, verify that $\forall \mathbf{x}\in \mathcal{C} \cap \mathcal{D}$, there exists $\mathbf{u}$ such that 
    \begin{equation}
    \label{eq:verification_formulation}
        \begin{bmatrix} 
            L_fV(\mathbf{x})\\
            -L_fh(\mathbf{x})\\
            \mathbf{0}
        \end{bmatrix} 
        + 
        \begin{bmatrix} 
            L_gV(\mathbf{x})\\
            -L_gh(\mathbf{x})\\
            A
        \end{bmatrix} 
        \mathbf{u} \leq  
        \begin{bmatrix} 
            -\kappa_V (V(\mathbf{x}))\\
            \kappa_h( h(\mathbf{x}))\\
            \mathbf{c}
        \end{bmatrix},
    \end{equation}
\end{problem}
namely there exists admissible $\mathbf{u}\in\mathcal{U}=\{\mathbf{u}|A\mathbf{u}\le \mathbf{c}\}$ satisfying both the CBF condition \eqref{def:cbf} and the CLF condition \eqref{def:clf} simultaneously. The region $\mathcal{C}\cap\mathcal{D}$ is control-invariant by definition, hence certified to be safe and stabilizable. We call the functions $V(\mathbf{x})$ and $h(\mathbf{x})$ compatible CLF and CBF if they satisfy the conditions of Problem \ref{prob:verification}, with $\mathcal{C}\cap\mathcal{D}$ as the compatible region.

\subsubsection{Compatibility Verification}
Verifying the condition \eqref{eq:verification_formulation} is equivalent to verifying that there exists a solution $\mathbf{u}$ for the  inequality 
\begin{equation}
\label{eq:verif_problem}
     \Lambda(\mathbf{x}) \mathbf{u} \leq  \xi(\mathbf{x}), \ \forall \mathbf{x}\in \mathcal{C} \cap \mathcal{D},
\end{equation}
where
\begin{equation}
    \Lambda(\mathbf{x}) = 
    \begin{bmatrix} 
        L_gV(\mathbf{x})\\
        -L_gh(\mathbf{x})\\
        A
    \end{bmatrix}, 
    \mathbf{\xi}(\mathbf{x}) = 
    \begin{bmatrix} 
        -\kappa_V (V(\mathbf{x}))-L_fV(\mathbf{x})\\
        \kappa_h( h(\mathbf{x}))+L_fh(\mathbf{x})\\
        \mathbf{c}
    \end{bmatrix}. \nonumber
\end{equation}
Note that $\Lambda(\mathbf{x}) \in \mathbb{R}^{(p+2)\times n_u}$ and $\mathbf{\xi}(\mathbf{x}) \in \mathbb{R}^{p+2}$. We let $\Lambda_i(\mathbf{x})$  denote the $i$-th column of $\Lambda(\mathbf{x})$.

Based on Farkas' Lemma, we give an equivalent condition (Lemma \ref{lemma:verify_farkas}) for CLF-CBF compatibility.

\begin{lemma}
\label{lemma:verify_farkas}
The CBF $h(\mathbf{x})$ and CLF $V(\mathbf{x})$ are compatible if and only if the following set is empty:
    \begin{multline}
    \label{eq:verify_farkas}
        \{(\mathbf{x},\mathbf{z})\; |\;h(\mathbf{x})\geq 0, V(\mathbf{x}) \leq 1,\\ 
        \mathbf{z}^T\Lambda(\mathbf{x})= \mathbf{0}^T, \mathbf{\xi}(\mathbf{x})^T\mathbf{z} =-1, \mathbf{z}\geq\mathbf{0}\},
    \end{multline}
 where $\mathbf{z}\in\mathbb{R}^{p+2}$ and $(\mathbf{x},\mathbf{z})\in\mathbb{R}^{n_x+p+2}$ is a concatenated vector.
\end{lemma}


Lemma \ref{lemma:verify_farkas} can be further simplified. Define the element-wise squared operation of vector $\mathbf{y} = [y_1,\dots, y_{p+2}]^T$ as $\mathbf{y}^2 = [y_1^2,\dots, y_{p+2}^2]^T$. Then an equivalent formulation of Lemma \ref{lemma:verify_farkas} is as follows: 
\begin{lemma}
\label{lemma:verify_farkas_symplify}
The CBF $h(\mathbf{x})$ and CLF $V(\mathbf{x})$ are compatible if and only if the following set is empty:
    \begin{multline}
    \label{eq:verify_farkas_simplify}
        \{(\mathbf{x},\mathbf{y})\;|\; h(\mathbf{x})\geq 0, V(\mathbf{x}) \leq 1,\\ 
        (\mathbf{y}^2)^T\Lambda(\mathbf{x})= \mathbf{0}^T, \mathbf{\xi}(\mathbf{x})^T\mathbf{y}^2 = -1\},
    \end{multline}
 where $\mathbf{y}^2\in\mathbb{R}_{\geq0}^{p+2}$ and $(\mathbf{x},\mathbf{y})\in\mathbb{R}^{n_x + p +2}$ is a concatenated vector. Namely, we replace the vector $\mathbf{z}$ and the non-negative constraint $\mathbf{z}\ge 0$ with a new vector $\mathbf{y}^2$ which is non-negative by construction.
\end{lemma}

Applying P-satz to Lemma \ref{lemma:verify_farkas_symplify}  gives  the following theorem. Noted that $(\mathbf{y}^2)^T\Lambda(\mathbf{x}) \in \mathbb{R}^{n_u}$, each element of $(\mathbf{y}^2)^T\Lambda(\mathbf{x})$ is computed by $(\mathbf{y}^2)^T\Lambda_i(\mathbf{x}), i=1, 2\dots n_u$ , and also, $\mathbf{\xi}(\mathbf{x})^T\mathbf{y}^2$ is a scalar. 

\begin{theorem}
\label{th:verify_p-satz}
The given CLF $V(\mathbf{x})$ and CBF $h(\mathbf{x})$ are compatible if and only if there exist polynomials $\lambda(\mathbf{x}, \mathbf{y}), \theta(\mathbf{x}, \mathbf{y})$ satisfying:
\begin{enumerate}
    \item $\theta(\mathbf{x}, \mathbf{y}) \in \Sigma[h(\mathbf{x}), 1- V(\mathbf{x})]$
    \item $\lambda(\mathbf{x}, \mathbf{y}) \in \mathbb{I}[ (\mathbf{y}^2)^T\Lambda_1(\mathbf{x}),\dots ,(\mathbf{y}^2)^T\Lambda_{n_u}(\mathbf{x}), \mathbf{\xi}(\mathbf{x})^T\mathbf{y}^2+1]$
    \item $\theta(\mathbf{x}, \mathbf{y}) + \lambda(\mathbf{x}, \mathbf{y}) + 1 = 0$
\end{enumerate}
\end{theorem}


As explained in section \ref{sec:mathematical_background}, directly applying P-satz might be computationally challenging. To remedy this, we could apply the \textit{S-procedure} (Lemma \ref{lemma:s-procedure}) and formulate the following convex optimization problem to verify compatibility
\begin{subequations}
\begin{align}
\text{Find}&\; \mathbf{s}_0(\mathbf{x}, \mathbf{y}), s_1(\mathbf{x}, \mathbf{y}), s_2(\mathbf{x}, \mathbf{y}), s_3(\mathbf{x}, \mathbf{y})\\
\begin{split}
\text{s.t }& -1 - \mathbf{s}_0(\mathbf{x}, \mathbf{y})^T((\mathbf{y}^2)^T\Lambda(\mathbf{x})) - s_1(\mathbf{x}, \mathbf{y})(\mathbf{\xi}(\mathbf{x})^T\mathbf{y}^2+1)\\&\qquad - s_2(\mathbf{x}, \mathbf{y})(1-V(\mathbf{x})) - s_3(\mathbf{x}, \mathbf{y})h(\mathbf{x}) \in \text{SOS}
\end{split}\label{eq:verify_S-procedure_constraint}\\
&s_2(\mathbf{x}, \mathbf{y}), s_3(\mathbf{x}, \mathbf{y}) \in \text{SOS}.
\end{align}
\label{eq:verify_S-procedure}
\end{subequations}

A given pair of CLF $V(\mathbf{x})$ and CBF $h(\mathbf{x})$ are compatible if we can find the polynomials $s_i(\mathbf{x},\mathbf{y}), i=0,1,2, 3$ in program \eqref{eq:verify_S-procedure}. 

If the system dynamics also include the algebraic constraint $e(\mathbf{x}) = 0$ (namely \eqref{eq:system_algebraic_constraint}), then we modify $\lambda(\mathbf{x}, \mathbf{y})$ in Theorem \ref{th:verify_p-satz} , by appending $e(\mathbf{x})$ to the ideal as $\lambda(\mathbf{x}, \mathbf{y}) \in \mathbb{I}[(\mathbf{y}^2)^T\Lambda_1(\mathbf{x}),\dots ,(\mathbf{y}^2)^T\Lambda_{n_u}(\mathbf{x}), \mathbf{\xi}(\mathbf{x})^T\mathbf{y}^2+1, e(\mathbf{x})]$, and we change the S-procedure in \eqref{eq:verify_S-procedure} accordingly. For simplicity, we will ignore the system algebraic constraint $e(\mathbf{x})=0$ in the remaining of this methodology section, and we will provide concrete examples with such constraints in Section \ref{sec: results} for better understanding.


\subsubsection{CBF Correctness Verification}
In addition to verifying the CLF-CBF compatibility, we must also ensure that the set $\{\mathbf{x}\;|\; h(\mathbf{x}) \geq 0\}$ is contained in the desired safe region, i.e., $\{\mathbf{x}\;|\;h(\mathbf{x})\geq 0\}$ implies $\mathbf{x} \in \mathcal{X}_s$.

Assume an unsafe region presented as:
\begin{equation}
\label{eq: unsafe_region_definition1}
    \mathcal{X}_u = \cup_{j=1}^{K}\{\mathbf{x}|l_j(\mathbf{x}) \le 0\}.
\end{equation} 
Using the S-procedure, we can  formulate the correctness verification program:
\begin{align}
    \text{Find } &q_1(\mathbf{x})\dots q_K(\mathbf{x}), p_1(\mathbf{x})\dots p_K(\mathbf{x}) \in \text{SOS}\nonumber\\
    \text{s.t. } & -1 - q_j(\mathbf{x})h(\mathbf{x}) + p_j(\mathbf{x})l_j(\mathbf{x}) \in \text{SOS},\;j = 1,\dots K.
    \label{eq:cbf_safe_constraint}
\end{align}
On the other hand, if the unsafe region is defined as
\begin{equation}
\label{eq: unsafe_region_definition2}
    \mathcal{X}_u = \cap_{j=1}^{K}\{\mathbf{x}|l_j(\mathbf{x}) \le 0\},
\end{equation}
then the correctness verification program is given by 
\begin{align}
    \text{Find } &q(\mathbf{x}), p_1(\mathbf{x})\dots p_K(\mathbf{x}) \in \text{SOS}\nonumber\\
    \text{s.t. } & -1 - q(\mathbf{x})h(\mathbf{x}) + \sum_{j=1}^{K}p_j(\mathbf{x})l_j(\mathbf{x})
\in \text{SOS}.
\label{eq:verify_correct2} 
\end{align}

Both problems \eqref{eq:cbf_safe_constraint} and \eqref{eq:verify_correct2} can be solved via SOS optimization.
\subsection{Compatible CBF and CLF synthesis}
\label{subsec: synthesis}
 Based on the previous verification framework, we next propose an alternating method to synthesize compatible CBF and CLF. For simplicity, in the following derivations, we use \eqref{eq: unsafe_region_definition1} as the definition of unsafe region and use the constraints in \eqref{eq:cbf_safe_constraint} as CBF correctness constraint.

As mentioned previously in the end of Section \ref{sec: prelim}, we set the kappa function $\kappa_V(\cdot)$ and $\kappa_h(\cdot)$ as linear functions with positive scalar constant coefficient, and we synthesize CLF/CBF with exponential stability/safety. This choice of linear $\kappa$ functions renders \eqref{eq:verify_S-procedure_constraint} to be a \textit{linear} function of the CLF/CBF, allowing us to search for CLF/CBF through convex optimization.
\begin{problem}
\label{prob:synthesis}
Given a system (\ref{eq:sys_dyn}), with equilibrium state $\mathbf{x}^*=\mathbf{0}$, constants $\kappa_V$, $\kappa_h$, and polynomials $l_1(\mathbf{x})\dots l_K(\mathbf{x})$ characterizing the unsafe region, find a positive definite function $V(\mathbf{x})$ and a function $h(\mathbf{x})$ such that there always exist a $\mathbf{u}\in\mathcal{U}$ satisfying \eqref{eq:verification_formulation} for all the $\mathbf{x}\in \mathcal{C}\cap \mathcal{D}$, where $\mathcal{C} = \{\mathbf{x}\;|\; h(\mathbf{x})\geq 0\}$, and $\mathcal{D} = \{\mathbf{x}\;|\; V(\mathbf{x}) \leq 1\}$.  We call the region $\mathcal{C}\cap \mathcal{D}$ as the ``compatible region", and our goal is to maximize some objective function $\gamma(\mathcal{C}\cap\mathcal{D})$ to expand this compatible region.
\end{problem}

Mathematically, our goal is to solve the following optimization problem
\begin{subequations}
\begin{align}
    &\max_{\substack{V(\mathbf{x}), h(\mathbf{x}), \\p(\mathbf{x}), q(\mathbf{x}), s(\mathbf{x})}} \gamma(\mathcal{C}\cap\mathcal{D})\\
    \text{s.t } &V(\mathbf{0}) = 0, V(\mathbf{x})\in\text{ SOS}\\
    &\text{Constraint } \eqref{eq:verify_S-procedure_constraint}, \eqref{eq:cbf_safe_constraint}\\
    &s_2(\mathbf{x}, \mathbf{y}), s_3(\mathbf{x}, \mathbf{y}), q_i(\mathbf{x}), p_i(\mathbf{x})\in \text{ SOS}, i=1,\hdots, K.
\end{align}
\label{eq:synthesize_bilinear}
\end{subequations}
Unlike the verification problem \eqref{eq:verify_S-procedure} and \eqref{eq:cbf_safe_constraint} where $V(\mathbf{x})$ and $h(\mathbf{x})$ are given, now we need to jointly search for $V(\mathbf{x}), h(\mathbf{x})$ together with the Lagrangian multipliers $p(\mathbf{x}), q(\mathbf{x}), s(\mathbf{x})$ in \eqref{eq:synthesize_bilinear}. Notice that all the constraints in problem \eqref{eq:synthesize_bilinear} are convex, except for \eqref{eq:verify_S-procedure_constraint} and \eqref{eq:cbf_safe_constraint}, which are \textit{bi-linear} in the decision variable $V(\mathbf{x}), h(\mathbf{x}), p(\mathbf{x}), q(\mathbf{x}), s(\mathbf{x})$, and hence cannot be directly handled by convex optimization. To remedy this, we resort to \textit{bilinear alternation}, a common technique in SOS-based controller design \cite{parrilo2000structured, tedrake2010lqr}. Where we alternate between
\begin{enumerate}
    \item Lagrangian searching: fix $V(\mathbf{x}), h(\mathbf{x})$, search for the Lagrangian multipliers $p(\mathbf{x}), q(\mathbf{x}), s(\mathbf{x},\mathbf{y})$.
    \item Update CLF-CBF: fix the Lagrangian multiplier $q(\mathbf{x}), s(\mathbf{x},\mathbf{y})$, search for $V(\mathbf{x}), h(\mathbf{x})$ and $p(\mathbf{x})$ while maximizing a surrogate function $\gamma(\mathcal{C}\cap\mathcal{D})$.
\end{enumerate}

Step 1 is the verification problem \eqref{eq:verify_S-procedure} and \eqref{eq:cbf_safe_constraint} discussed in the previous subsection. In step 2, we consider the following objective to expand the compatible region $\mathcal{C}\cap\mathcal{D}$: 
 given a set of predefined ``candidate states" $\mathbf{x}^{(1)}, \hdots, \mathbf{x}^{(L)}$, we want the compatible region $\mathcal{C}\cap\mathcal{D}$ to cover as many ``candidate states" as possible. We know that the candidate state $\mathbf{x}^{(i)}\in\mathcal{C}\cap\mathcal{D}$ iff $V(\mathbf{x}^{(i)})\le 1$ and $h(\mathbf{x}^{(i)}) \ge 0$. Hence we can design the objective function as minimizing
\begin{align}
     \sum_{i=1}^Lc_1\text{ReLU}\left(V(\mathbf{x}^{(i)})-1\right) + c_2\text{ReLU}\left(-h(\mathbf{x}^{(i)})\right)\label{eq:candidate_states_cost},
\end{align}
where $c_1, c_2 > 0$ are given positive weights, and Eq \eqref{eq:candidate_states_cost} penalizes the violation $x^{(i)}\notin \mathcal{C}\cap\mathcal{D}$. To avoid arbitrary scaling of CBF $h(\mathbf{x})$ (if $h(\mathbf{x})$ is a valid CBF, the scaling it as $k*h(\mathbf{x})$ by any positive constant $k$ is still a valid CBF, and the cost function becomes unbounded under infinite scaling), we impose an additional constraint $h_{\text{lower}}\le h(\mathbf{0}) \le h_{\text{upper}}$, where $h_{\text{lower}}, h_{\text{upper}}$ are given positive constants. As a result, the optimization problem is written as
\begin{subequations}
    \begin{align}
        \min_{V(\mathbf{x}), h(\mathbf{x}), p(\mathbf{x})} &\text{Objective }\eqref{eq:candidate_states_cost}\\
        \text{s.t }& V(\mathbf{0})=0, V(\mathbf{x})\in \text{ SOS}\\
        &\text{Constraint } \eqref{eq:verify_S-procedure_constraint}, \eqref{eq:cbf_safe_constraint}\\
        &h_{\text{lower}}\le h(\mathbf{0})\le h_{\text{upper}}.
    \end{align}
    \label{eq:cover_candidate_states}
\end{subequations}
The optimization problem \eqref{eq:cover_candidate_states} is a convex optimization problem when searching over $V(\mathbf{x}), h(\mathbf{x})$ and $p(\mathbf{x})$.

To summarize, we outline our synthesis algorithm in Algorithm \ref{alg:synthesis}.
\begin{algorithm}
\caption{Synthesize compatible CLF-CBF}
\label{alg:synthesis}
\begin{algorithmic}[1]
\State Given system dynamics \eqref{eq:sys_dyn}, positive constants $\kappa_V, \kappa_h$ and unsafe region $\mathcal{X}_u$ in \eqref{eq: unsafe_region_definition1}.
\State Initialize $V^{(0)}(\mathbf{x}), h^{(0)}(\mathbf{x})$. $i=0$
\While{$i < \text{MaxIter}$}
\State Fix $V^{(i)}(\mathbf{x}), h^{(i)}(\mathbf{x})$, solve \eqref{eq:verify_S-procedure} and \eqref{eq:cbf_safe_constraint} to find the Lagrangian multipliers $p^{(i)}(\mathbf{x}), q^{(i)}(\mathbf{x}), s^{(i)}(\mathbf{x}, \mathbf{y})$.
\State Fix $q^{(i)}(\mathbf{x}), s^{(i)}(\mathbf{x}, \mathbf{y})$, find CLF-CBF $V^{(i+1)}(\mathbf{x}), h^{(i+1)}(\mathbf{x})$ and Lagrangian $p^{(i+1)}(\mathbf{x})$ through solving \eqref{eq:cover_candidate_states}.
\State $i\leftarrow i+1$.
\EndWhile
\end{algorithmic}
\end{algorithm}
In line 2 of Algorithm \ref{alg:synthesis}, we can initialize the CLF/CBF in many ways, for example a good candidate is to linearize the dynamics around the equilibrium state and use the LQR solution to design the initial guess (the LQR solution with the equality constraint $e(\mathbf{x})=0$ can be computed by projecting the dynamics onto the manifold $e(\mathbf{x})=0$, see \cite[chapter~8.3.2]{underactuated}).

With a feasible initial guess $V^{(0)}$ and $h^{(0)}$, by induction, the CLF $V^{(i)}$ and CBF $h^{(i)}$ are also compatible in each iteration.

\section{Results}
\label{sec: results}
In this section, we apply our algorithm to different dynamical systems. We choose a toy system whose results can be easily visualized on 2D plots, and a 13-state quadrotor robot to showcase the scalability of our approach. All optimization problems are solved via Mosek solver \cite{mosek}. We report the CLF/CBF and Lagrangian degrees, together with the computation time in Table \ref{table:degree_and_time}.
\begin{table*}[t]
\centering
\begin{tabular}{|c|c|c|c|c|c|c|c|c|c|c|}
\hline
& \multicolumn{8}{c|}{Polynomial degree} & \multicolumn{2}{c|}{Computation time}\\
\hline
&$V(\mathbf{x})$&$h(\mathbf{x})$&$s_0(\mathbf{x},\mathbf{y})$ & $s_1(\mathbf{x},\mathbf{y})$ & $s_2(\mathbf{x},\mathbf{y})$ & $s_3(\mathbf{x},\mathbf{y})$ & $p(\mathbf{x})$ & $q(\mathbf{x})$ & Problem \eqref{eq:verify_S-procedure} & Problem \eqref{eq:cover_candidate_states}\\
\hline
Toy & 2& 2 & 2 & 2 & 4 & 4 & 0 & 0& 0.05s& 0.08s\\
\hline
Quadrotor & 2& 2& 1 & 1 & 4 & 4 & 0 & 0 & 213s&149s\\
\hline
\end{tabular}
\caption{Degrees of CLF/CBF, Lagrangian polynomials and computation time.}
\label{table:degree_and_time}
\end{table*}

\subsection{Toy system}
\label{subsec: toy_system}
We consider the following toy system (from \cite{tan2004searching})
\begin{subequations}
\begin{align}
    \dot{\theta} =& u\\
    \dot{\gamma} =& -\sin\theta - u\\
    -1 \le &u \le 1,
\end{align}
\end{subequations}
where the unsafe region is
\begin{align}
    \mathcal{X}_u = \{(\theta, \gamma) | \sin\theta + \cos\theta + \gamma + 1 \le 0\}.
\end{align}
First, we convert the system dynamics to the polynomial form, by introducing new variables $s = \sin\theta, c=\cos\theta$, with the state defined as $\mathbf{x}=[x_1, x_2, x_3] = [s, c-1, \gamma]$. Note that we subtract $1$ from $\cos\theta$ to make sure that the equilibrium $\theta=0,\gamma=0$ is at $\mathbf{x}=\mathbf{0}$. The polynomial dynamics is
\begin{subequations}
\begin{align}
    \dot{\mathbf{x}} = \begin{bmatrix}0\\0\\-x_1\end{bmatrix} + \begin{bmatrix} x_2+1\\-x_1\\-1\end{bmatrix}u\\
    -1\le u \le 1,
\end{align}
with the constraint $\sin^2\theta + \cos^2\theta=1$ as the following algebraic constraint on the state $\mathbf{x}$
\begin{align}
    e(\mathbf{x}) = x_1^2 + (x_2+1)^2 -1 = 0.
\end{align}
\end{subequations}
The unsafe region is $\mathcal{X}_u = \{\mathbf{x} | x_1 + x_2 + x_3 + 2 \le 0\}$.

\begin{figure}
\begin{subfigure}{0.48\textwidth}
\includegraphics[width=0.98\textwidth]{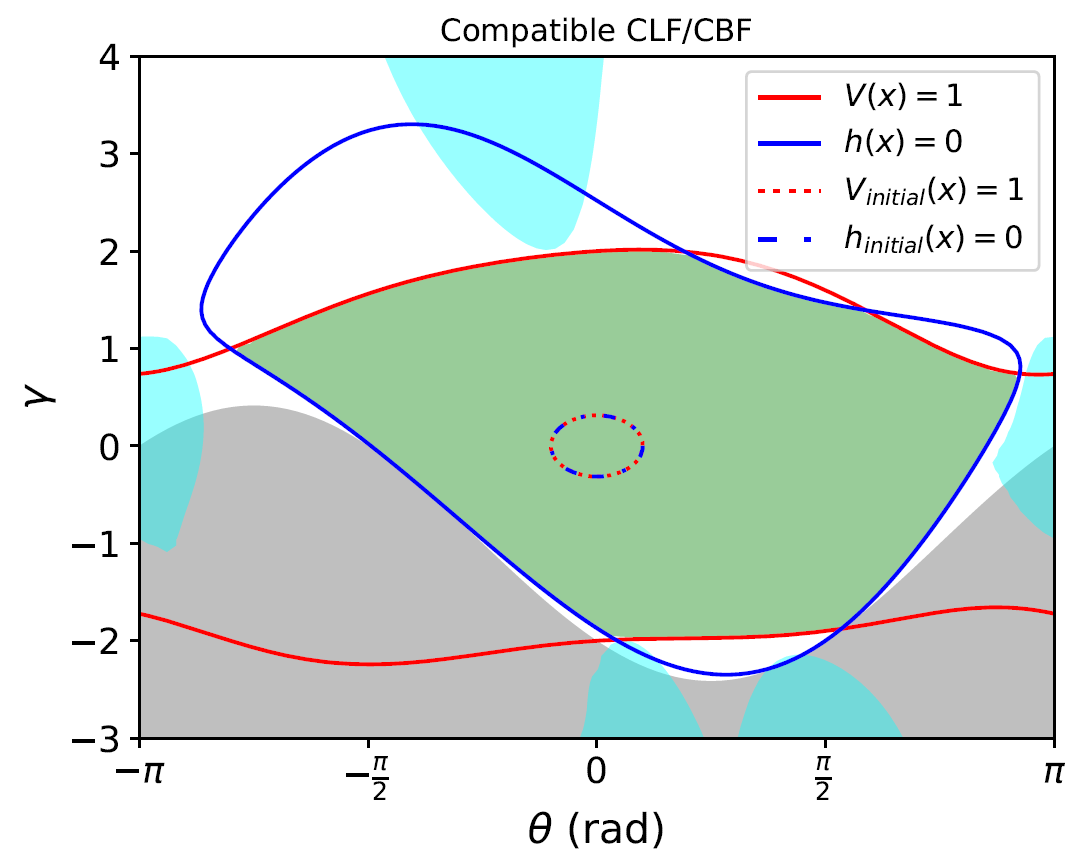}
\subcaption{Synthesize compatible CLF/CBF. The grey region is the obstacle. The cyan region is where CLF is incompatible with the CBF. The compatible region grows from the small circle $\{\mathbf{x}|V_{\text{initial}}(\mathbf{x})\le1, h_{\text{initial}}(\mathbf{x})\ge 0\}$ to the large green region. The green compatible region doesn't touch the grey obstacle or the cyan incompatible region.}
\label{fig:nonlinear_toy_compatible}
\end{subfigure}
\begin{subfigure}{0.48\textwidth}
\includegraphics[width=0.99\textwidth]{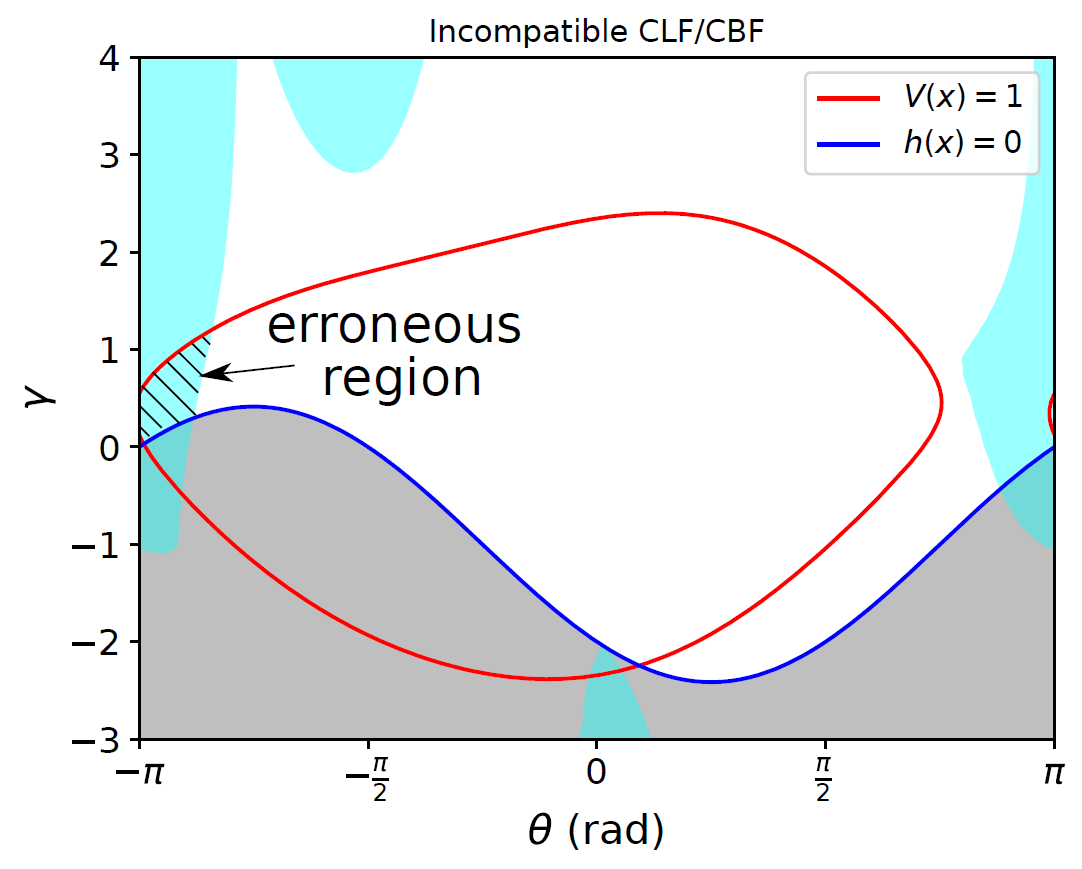}
\subcaption{Synthesize CLF and CBF separately using approaches from previous works. They are incompatible with each other, as demonstrated in the highlighted erroneous region, where the set $\{\mathbf{x} | V(\mathbf{x})\le 1, h(\mathbf{x})\ge 0\}$ intersects with the cyan incompatible region in which there doesn't exist a control action to satisfy CLF constraint and CBF constraint simultaneously.}
\label{fig:nonlinear_toy_incompatible}
\end{subfigure}
\caption{Results of jointly synthesizing compatible CLF/CBF (Fig \ref{fig:nonlinear_toy_compatible}, our approach), and separately synthesizing incompatible CLF/CBF (Fig \ref{fig:nonlinear_toy_incompatible}, previous approach) for the nonlinear toy example.}
\end{figure}

We synthesize compatible CLF/CBF through algorithm \ref{alg:synthesis}. We choose the initial CLF as $V_{\text{initial}}(\mathbf{x}) = 10(x_1^2 + x_2^2 + x_3^2)$ and the initial CBF as $h_{\text{initial}}(\mathbf{x}) = 1 - V_{\text{initial}}(\mathbf{x})$. We plot the result in Fig.\ref{fig:nonlinear_toy_compatible} (note that the plot is in the original $(\theta, \gamma)$ space). We also draw the \textit{incompatible region}, namely the region where the CLF constraint $L_fV(\mathbf{x}) + L_gV(\mathbf{x})u + \kappa_VV(\mathbf{x})\le 0$ and the CBF constraint $L_fh(\mathbf{x}) + L_gh(\mathbf{x})u + \kappa_hh(\mathbf{x})\ge 0$ don't share a common solution for $u$ within the input limit $u\in [-1, 1]$. Fig \ref{fig:nonlinear_toy_compatible} shows that
\begin{enumerate}
    \item Algorithm \ref{alg:synthesis} can successfully find compatible CLF and CBF that expand the compatible region $\{\mathbf{x}|V(\mathbf{x})\le 1, h(\mathbf{x})\ge 0\}$, which grows from the small ellipsoid in the center to the large green region.
    \item The synthesized compatible region is very tight, as the green compatible region almost touches the grey obstacle region and the cyan incompatible region. This showcases that our S-procedure formulation \eqref{eq:verify_S-procedure} can generate very tight solution with low-degree Lagrangian multipliers (the Lagrangians are of degree no larger than 4).
\end{enumerate}

As a comparison, we use the previous approach \cite{dai2023convex} to synthesize the CLF and CBF separately (The CBF is set as $h(\mathbf{x}) = x_1 + x_2 + x_3+2$, such that $\{\mathbf{x}\;|\;h(\mathbf{x}) \ge 0\}$ is exactly the safe region $\mathcal{X}_s$). As shown in Fig \ref{fig:nonlinear_toy_incompatible}, ignoring the compatibility condition during CLF/CBF synthesis leads to the existence of the erroneous region, in which the CLF constraint and CBF constraint cannot be satisfied simultaneously, causing the CLF-CBF-QP controller to fail at the runtime.

\subsection{Quadrotor}
To showcase the scalability of our approach, we apply it to a 3D quadrotor with 13 states, including quaternion for its body orientation, the position, linear velocity and angular velocity. The input $\mathbf{u}\in\mathbb{R}^4$ is the rotor thrust. We have the extra constraint $\mathbf{q}^T\mathbf{q}=1$ on the unit-length quaternion $\mathbf{q}$. The polynomial dynamics of the quadrotor is described in \cite{Fresk2013}. The obstacle is the ground at height $-0.5m$. 

\begin{figure}
    \includegraphics[width=0.45\textwidth]{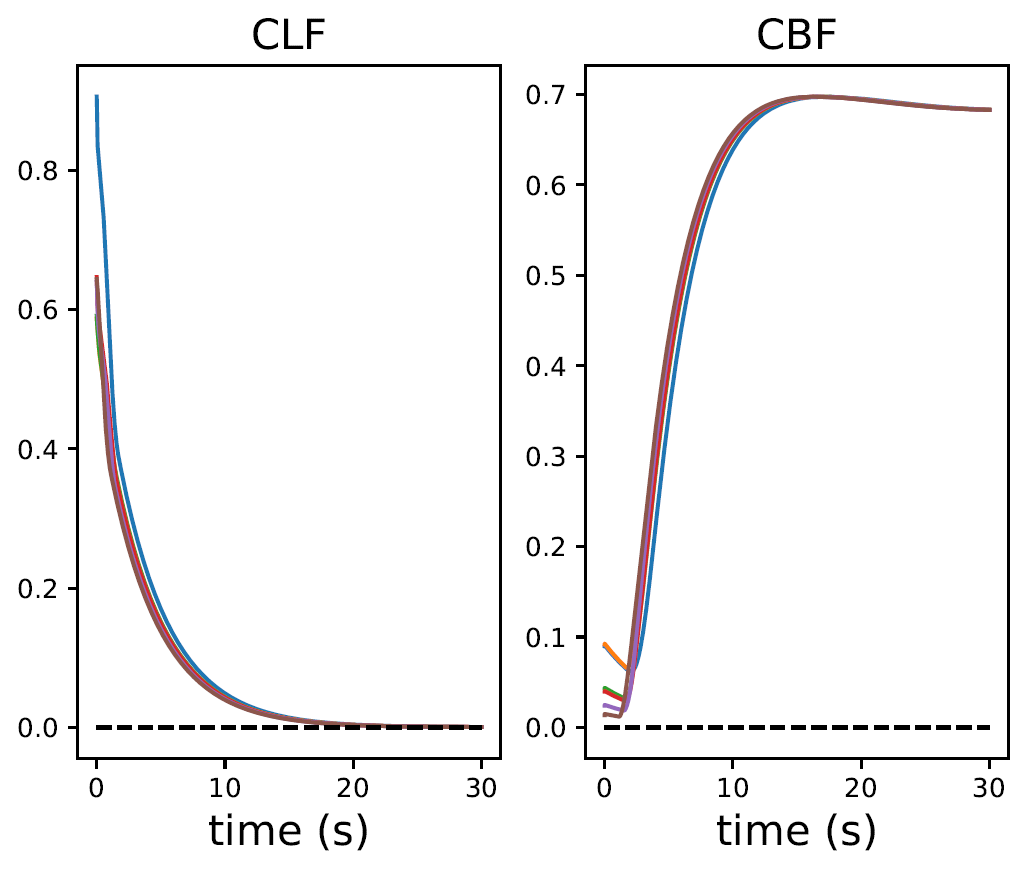}
    \caption{CLF/CBF trajectories of the quadrotor while simulated with the CLF-CBF-QP controller.}
    \label{fig:quadrotor_clf_cbf}
\end{figure}

We synthesize the compatible CLF/CBF through Algorithm \ref{alg:synthesis}\footnote{For better numerics, we use SDSOS constraint instead of SOS constraint on the Lagrangian multiplier $s_2(\mathbf{x},\mathbf{y}), s_3(\mathbf{x},\mathbf{y})$, where SDSOS is a stronger convex constraint than SOS \cite{ahmadi2019dsos}.}, with the initial CLF computed from the LQR solution, and the initial CBF as $1-V_{\text{initial}}(\mathbf{x})$. Note that the CBF depends on both the position and velocity, hence the relative-degree of the CBF equals to 1. After optimizing the CLF/CBF, we simulate the quadrotor with the CLF-CBF-QP controller, whose cost is to minimize the thrust $\mathbf{u}^T\mathbf{u}$. We start the simulation from different initial states within the compatible region, as shown in Fig \ref{fig:quadrotors}. Since the QP objective is to minimize the thrust, the quadrotor first descends (with small thrusts), and then the CBF constraint will prevent it from hitting the ground and the quadrotor lifts up to the goal. We also draw the CLF/CBF trajectories along the simulation in Fig \ref{fig:quadrotor_clf_cbf}. The CLF values converge to 0 while the CBF values stay above 0.
\section{Conclusion}
\label{sec: conclusion}
In this paper, for control affine systems both with and without input limits, we studied the verification and synthesis of compatible CBF and CLF independent from any nominal controllers. We presented  necessary and sufficient conditions for compatible CBF and CLF via Farkas' Lemma and algebraic-geometric Positivstellensatz, and demonstrated that these conditions can be verified by solving a SOS program. In order to reduce the number of sos polynomials, we further simplified the necessary and sufficient condition using S-procedure and formulated a more tractable SOS program for compatibility verification. 
We solve a sequence of SOS programs so as to synthesize CLF/CBF with a large compatible region. Finally, we evaluated our proposed approach on a toy system and a quadrotor. 

In our future work, it is natural to extend the verification and synthesis framework to high relative degree CBF \cite{xiao2019control}. Moreover, in this work we used a single polynomial function as the CBF. It would require a very high degree polynomial to capture the safe region with a complicated shape, resulting in a large-size optimization problem and poor numerics. To remedy this, we will consider using the minimal of several compatible low-degree polynomials as the CBF, which we intend to address in our follow-up work.

\bibliographystyle{ieeetr}
\bibliography{ref}
\end{document}